\documentclass[twocolumn,aps,prl,amsmath,showpacs,amssymb,floatfix]{revtex4}
\usepackage{tabularx}
\usepackage{bm}
\usepackage{subfigure}
\usepackage{euscript}
\usepackage{graphicx}
\usepackage{color}
\usepackage[colorlinks=true,linkcolor=blue]{hyperref}%
\usepackage{amsfonts}
\usepackage{exscale}
\usepackage{amsbsy}

\begin{document}

\title{Floquet Quantum Spin Hall Insulator in Cold Atomic Systems}

\author{Zhongbo Yan$^1$}
\author{Bo Li$^{1}$}
\author{Xiaosen Yang$^{2}$}
\author{Shaolong Wan$^1$}
\email[]{slwan@ustc.edu.cn}
\affiliation{$^{1}$Institute for Theoretical
Physics and Department of Modern Physics University of Science and
Technology of China, Hefei, 230026, P. R. China\\
$^2$Beijing Computational Science Research Center, Beijing,
100084, P. R. China}
\date{\today}

\begin{abstract}
For cold atomic systems, varying the optical lattice potential
periodically provides a general and simple way to drive the system
into phases with nontrivial topology. Besides its simplicity, this
driving approach, compared to the usual driving approach by
exerting an external electromagnetic field to the static system,
has the merit that it does not break the original static system's
time-reversal symmetry at any given time. Based on this approach,
we find that a trivial insulator with time-reversal symmetry can
be driven into a Floquet quantum spin Hall insulator. This novel
state of matter can stably host one or two pair of gapless helical
states on the same boundary, which suggests this state is not a
simple analog of the quantum spin Hall insulator. The effect of a
time-reversal-symmetry-breaking periodic perturbation, the
stability of the novel states, and this new driving approach to a
system without time-reversal symmetry are discussed.
\end{abstract}

\pacs{71.10.Fd, 03.65.Vf, 73.43.-f}

\maketitle

{\it Introduction.---} In the past few years, the theoretical
predictions \cite{C. L. Kane1, C. L. Kane2, B. A. Bernevig} and
the experimental observations \cite{M. Konig, D. Hsieh} of
topological insulator have stimulated strong and continuous
interest in predicting new materials and systems with topological
phases due to their potential application in spintronic and
topological quantum computation \cite{A. Kitaev}.

Topological phases exist in every dimension \cite{A. P. Schnyder,
A. Y. Kitaev}, however, it is found that the number of real static
systems with topological properties is quite limited. This
limitation triggers new proposals to engineer systems with
topological properties. One such proposal that time-periodically
driven systems can host topological characteristics, the so-called
Floquet approach \cite{T. Kitagawa1, N. H. Lindner}, recently has
attracted great attention \cite{J.-I. Inoue,  Z. Gu, Ferenc Simon,
G. C. Liu, G. Platero, D. L. Bergman, A. A. Reynoso, Y. T. Katan,
D. E. Liu, A. Kundu, M. Lababidi, P. M. Perez-Piskunow, A. G.
Grushin}, and has been demonstrated by the direct observation of
protected edge modes in photonic crystals \cite{M. C. Rechtsman,
T. Kitagawa2}. The great interest arisen are not only this
approach can drive a topologically trivial system to be
topologically nontrivial but also the driven systems can exhibit
unique topological properties without an analog in static systems
\cite{T. Kitagawa3, M. S. Rudner}, such as the Floquet Majorana
fermions with quasienergy $\epsilon=\pi/T$ \cite{L. Jiang}.

For solid state systems, currently the general way to drive the
system from a trivial phase into a topological phase is  by
introducing a time periodic external electromagnetic field to the
original static system, like shining light on a conventional
insulator. This method is simple and easy to control, however, it
has a drawback that if the external field, like an oscillating
magnetic field, directly couples with the spin, it usually breaks
the time-reversal symmetry of a spinful fermionic system in the
sense:
$\mathcal{T}\hat{H}_{0}(-k)\mathcal{T}^{-1}=\hat{H}_{0}^{*}(k)$
but $\mathcal{T}\hat{H}(-k,t)\mathcal{T}^{-1}\neq\hat{H}^{*}(k,t)$
for any given $t$ at which the external field is nonzero, where
$\hat{H}_{0}(k)$ is the original static Hamiltonian and
$\hat{H}(k,t)$ is the driven Hamiltonian,
$\mathcal{T}=i\sigma_{y}\mathcal{K}$. The consequence of this
drawback is that if the external field only induces driving terms
that break the time-reversal symmetry, a topologically trivial
system with time-reversal symmetry can never be driven to be
topologically nontrivial. To avoid this drawback, therefore a
general driving method which guarantees inducing terms that do not
break the time-reversal symmetry is favored.

For cold atomic systems, we find there exists such a general and
simple method to drive the system periodically without breaking
the time-reversal symmetry if the original static system is
time-reversal invariant. The method is varying the optical
potential periodically, which means $V({\bf r},t)=V({\bf r},t+T)$.
A direct consequence of periodically varying the optical lattice
is that the hopping amplitude of a tight-binding model will turn
to be periodic. For a time-reversal invariant static system, the
hopping amplitude turning to be time-periodic does not break the
time-reversal symmetry. As a consequence, for a time-reversal
invariant insulator, the edge states driven up are always helical,
the same as the Quantum Spin Hall (QSH) insulator \cite{C. L.
Kane1, C. L. Kane2, B. A. Bernevig} (for a time-reversal invariant
superconductor or superfluid, the picture is similar and therefore
we restrict ourself to insulator in this work). For the sake of
accuracy, here we name systems hosting such driven-up helical edge
states as Floquet Quantum Spin Hall (FQSH) insulator.

{\it Theoretical model with time-reversal symmetry---} We consider
a cold atomic realization of the time-reversal symmetric Kane-Mele
model in a hexagonal optical lattice. The Hamiltonian is given by
\cite{C. L. Kane1}
\begin{eqnarray}
H_{KM}=J\sum_{<i,j>}c^{\dag}_{i}c_{j}+
i\lambda_{SO}\sum_{<<i,j>>}\nu_{ij}c^{\dag}_{i}\sigma_{z}c_{j}+
\lambda_{v}\sum_{i}\xi_{i}c^{\dag}_{i}c_{i}.\label{1}
\end{eqnarray}
The first term denotes the nearest-neighbor hopping process. The
second term is the mirror symmetric spin-orbit interaction which
involves spin-dependent next-nearest-neighbor hopping. $\nu_{ij}$
takes value $1$ (or $-1$) when the path $i\rightarrow j$ is
contourclockwise (or clockwise). The third term is a staggered
sublattice potential ($\xi_{i}=\pm$), which are included to
control the phase.

For a hexagonal lattice, the optical lattice potential takes the
form \cite{L.-M. Duan}
\begin{eqnarray}
V({\bf r},t)=\sum_{i=1,2,3}V_{i}(t)\sin^{2}[k_{L}(x\cos\theta_{i}+
y\sin\theta_{i})+\frac{\pi}{2}],\label{2}
\end{eqnarray}
where $\theta_{1}=\pi/3$, $\theta_{2}=2\pi/3$ and $\theta_{3}=0$.
$k_{L}$ is the optical wave vector. Here we consider a hexagonal
optical lattice with isotropic driving, $i.e.$
$V_{i}(t)=V_{0}+V_{D}\cos(\omega t)$. With such a driving, the
hopping amplitudes correspondingly vary with time periodically:
$J(t)=J+J_{D}\cos(\omega t)$,
$\lambda_{SO}(t)=\lambda_{SO}+\lambda_{SO,D}\cos(\omega t)$. Then
the time-dependent Hamiltonian can be decomposed as
$H(t)=H_{KM}+H_{D}\cos(\omega t)$. $H(t)$ is time-periodic and
$H_{D}$ is given by
\begin{eqnarray}
H_{D}=J_{D}\sum_{<i,j>}c^{\dag}_{i}c_{j}+i\lambda_{SO,D}\sum_{<<i,j>>}
\nu_{ij}c^{\dag}_{i}\sigma_{z}c_{j}.\label{3}
\end{eqnarray}
$H_{D}$ has the same form as $H_{KM}$ except the absence of a
corresponding term to $\lambda_{v}$ (in fact, even if we include
such a corresponding term, the conclusion is not affected), and
therefore does not break the time-reversal symmetry, and the total
Hamiltonian $H(t)$ still hold the time-reversal symmetry.

The single-particle Schr\"{o}dinger equation associated with this
time-dependent Hamiltonian is:
\begin{eqnarray}
[\mathcal{H}(\mathbf{k},t)-i\partial_{t}]\Psi(\mathbf{k},t)=0, \,
\text{with}\, \mathcal{H}(\mathbf{k},t+T) =
\mathcal{H}(\mathbf{k},t).\label{4}
\end{eqnarray}
where $\mathcal{H}(\mathbf{k},t)$ is the form of $H(t)$ in
momentum space. According to the Bloch-Floquet theory, the wave
function satisfying Eq.(\ref{4}) can be expressed as
$\Psi(\mathbf{k},t)=\text{e}^{-i\varepsilon_{\mathbf{k}}t}\Phi(\mathbf{k},t)$
with the Floquet states $\Phi(\mathbf{k},t+T)=\Phi(\mathbf{k},t)$
and the Floquet equation
$[\mathcal{H}(\mathbf{k},t)-i\partial_{t}]
\Phi(\mathbf{k},t)=\varepsilon_{\mathbf{k}}\Phi(\mathbf{k},t)$.
The parameter $\varepsilon$, called the quasienergy, is uniquely
defined up to integer multiples of $\omega=2\pi/T$. Similar to the
crystal momentum of a system with discrete translation symmetry,
the quasienergy can be thought of as a periodic variable defined
on a quasienergy Brillouin zone $-\pi/T<\varepsilon\leq\pi/T$.

Although there are many different (but equivalent) ways to compute
the topological invariant for a time-reversal symmetric insulator
\cite{Z. Wang}, to the best of our knowledge, a direct way to
calculate the topological invariant for a time-reversal symmetric
driven model is still lacked. To determine the topological
property of the time-dependent Hamiltonian, here we use the
`repeated zone analysis' \cite{M. S. Rudner}. The first step is to
expand the Floquet states,
$\Phi(\mathbf{k},t)=\sum_{m}\phi_{m}(\mathbf{k})\text{e}^{im\omega
t}$. The coeffcients $\phi_{m}(\mathbf{k})$ satisfy the
time-independent eigenvalue equation
\begin{eqnarray}
\sum_{m^{'}}\mathcal{H}_{m,m^{'}}(\mathbf{k})\phi_{m^{'}}(\mathbf{k})
=\varepsilon_{\mathbf{k}}\phi_{m}(\mathbf{k}),\label{5}
\end{eqnarray}
where the matrix form Floquet Hamiltonian $\mathcal{H}_{mm^{'}}(\mathbf{k})$ is given by
\begin{eqnarray}
\mathcal{H}_{m,m^{'}}(\mathbf{k})=m\omega\delta_{mm^{'}}+
\frac{1}{T}\int_{0}^{T}dt e^{-i(m-m^{'}) \omega t}
\mathcal{H}(\mathbf{k},t).\label{6}
\end{eqnarray}
Write more explicitly,
\begin{eqnarray}
&&\mathcal{H}_{m,m}(\mathbf{k})=m\omega+\mathcal{H}_{KM}(\mathbf{k}),\nonumber \\
&&\mathcal{H}_{m,m+1}(\mathbf{k})=\mathcal{H}_{m+1,m}(\mathbf{k})=\frac{1}{2}
\mathcal{H}_{D}(\mathbf{k}). \label{7}
\end{eqnarray}
The matrix $\mathcal{H}_{m,m^{'}}(\mathbf{k})$ has the block tridiagonal form,
where each block is a $2\times2$ matrix.

According to the bulk-edge correspondence, the absence or appearance of edge
states traversing the gaps reflects that the system is topologically trivial
or topologically nontrivial, respectively. To see whether the driven system
hosts edge states, we consider the system with periodical boundary condition
in $x$ direction and open boundary condition in $y$ direction (ZigZag geometry).

\begin{figure}
\subfigure{\includegraphics[width=4cm, height=4cm]{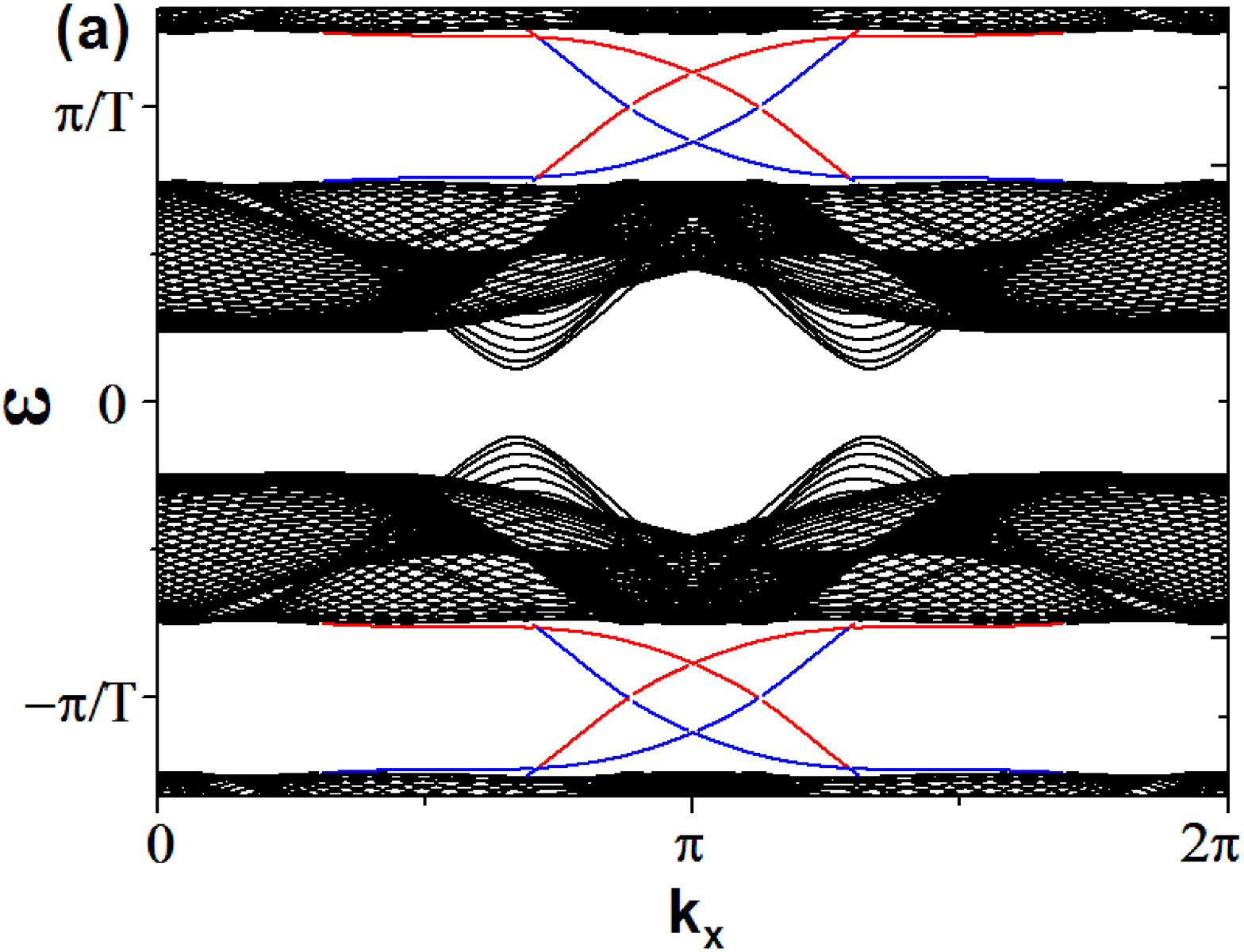}}
\subfigure{\includegraphics[width=4cm, height=4cm]{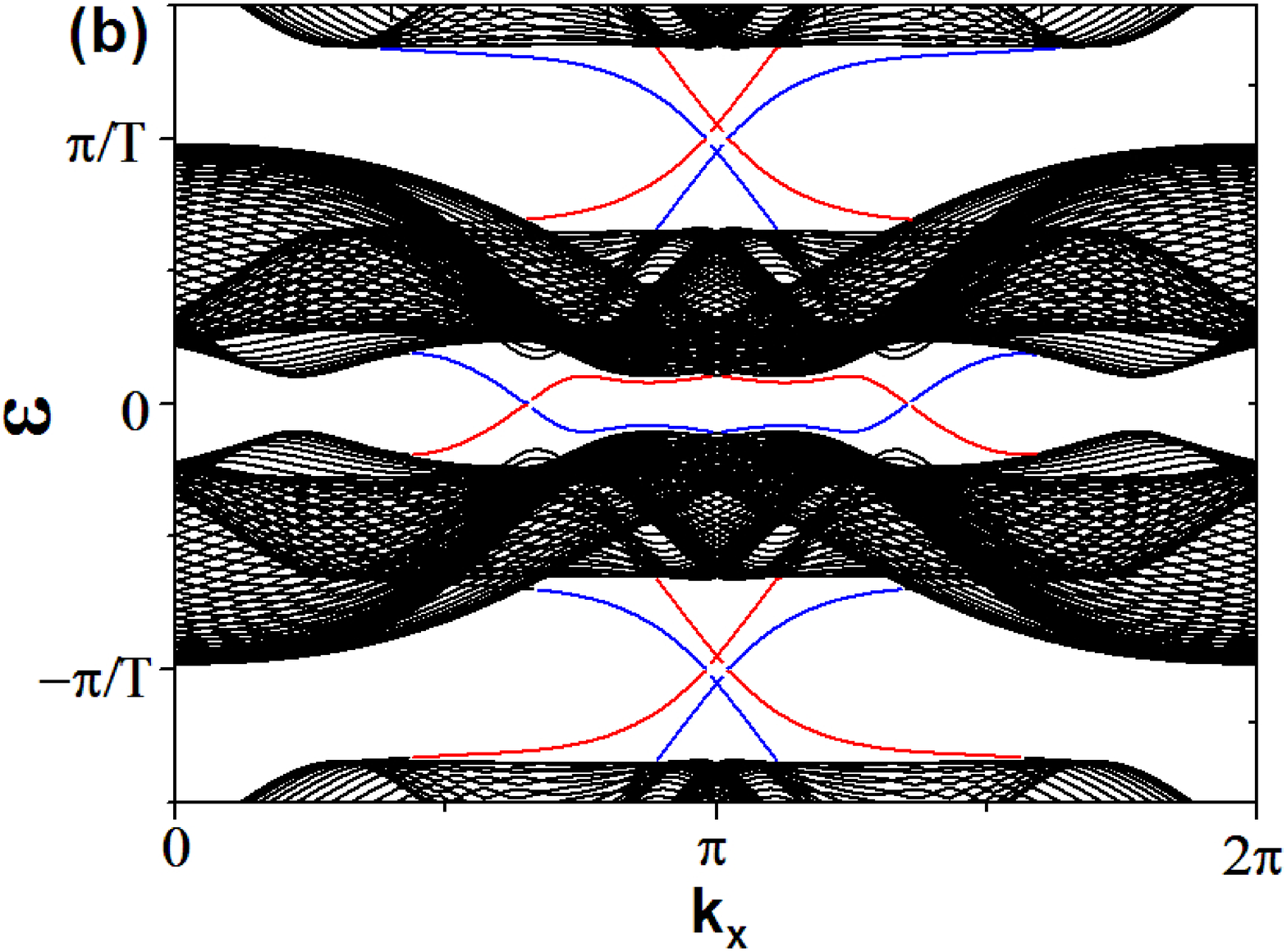}}
\subfigure{\includegraphics[width=4cm, height=4cm]{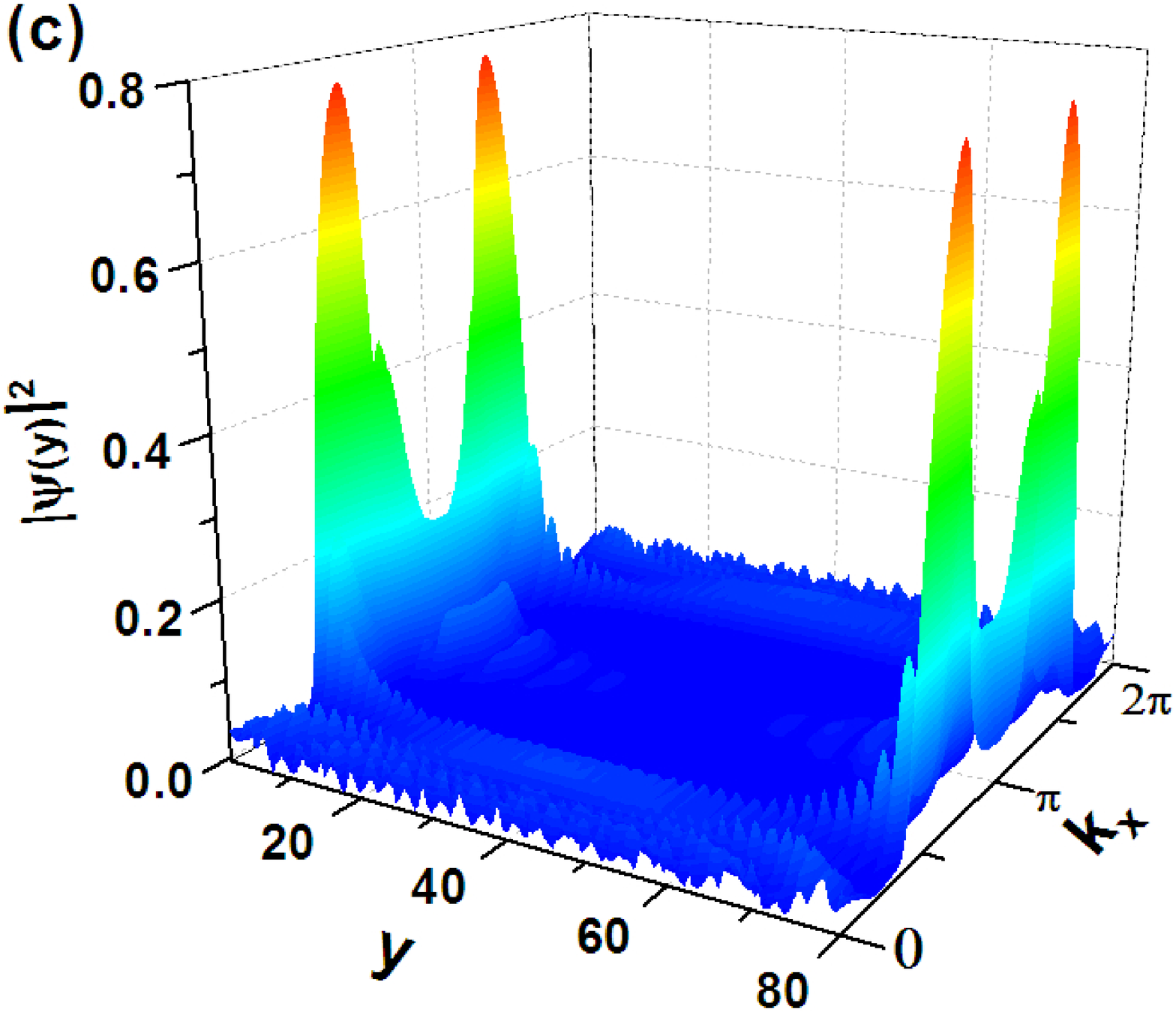}}
\subfigure{\includegraphics[width=4cm, height=4cm]{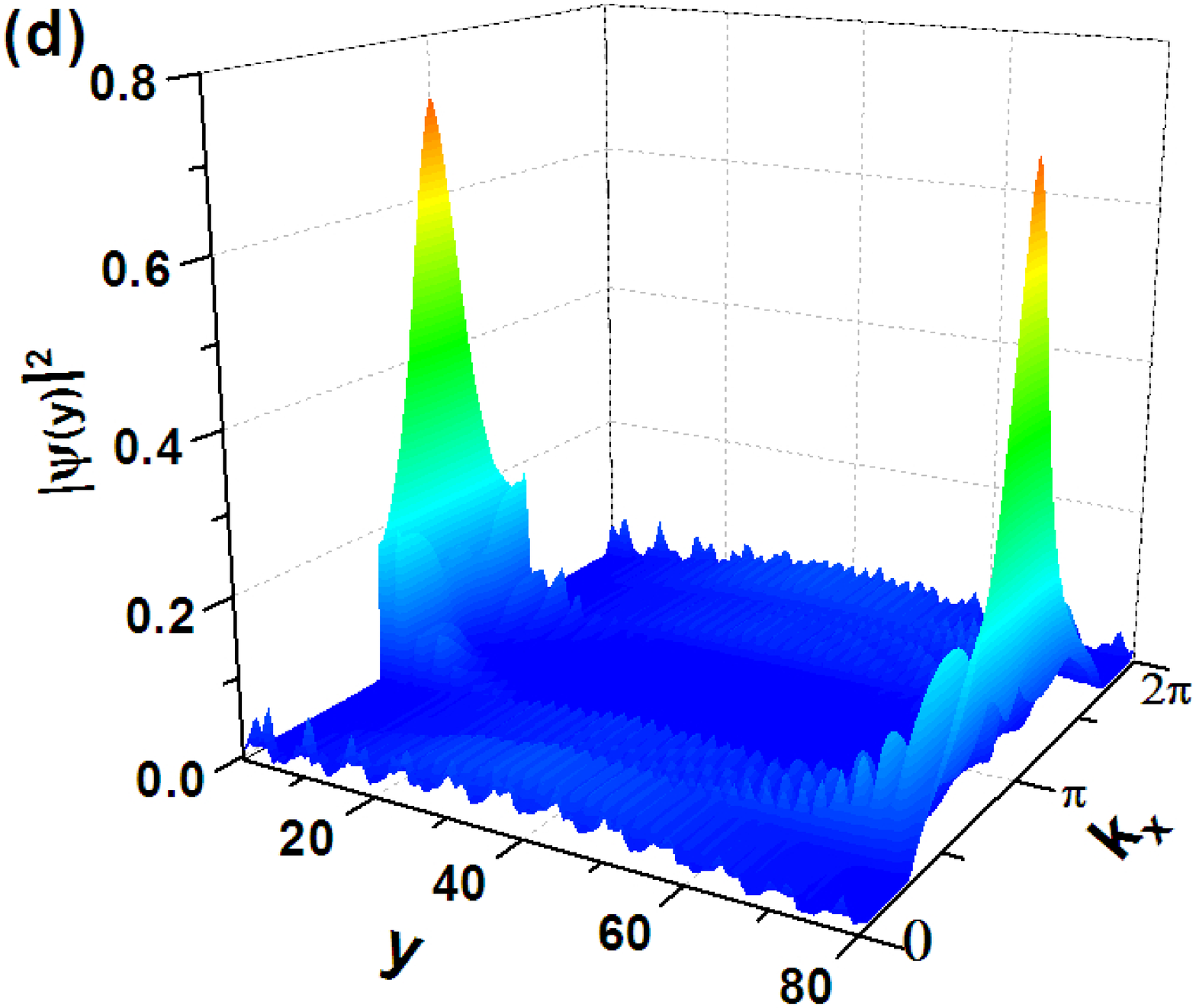}}
\caption{ (color online) (a)(b) Spectrum of the truncated Floquet
Hamiltonian (Eq.(\ref{6})) in the ZigZag geometry. The parameters
used in (a) are $J=1$ (energy unit), $\lambda_{SO}=0.2$,
$\lambda_{v}=1.2$, $J_{D}=1$, $\lambda_{SO,D}=0.1$, $\omega=3$.
The parameters in (b) are the same as (a) except now $\omega=2$.
(c) The density of the edge states traversing the gap at
$\varepsilon=0$. (d) The density of the edge states traversing the
gap at $\varepsilon=\pi/T$.}\label{fig1}
\end{figure}

In Fig.\ref{fig1}(a), the  static parameters are chosen as $J=1$,
$\lambda_{SO}=0.2$, $\lambda_{v}=1.2$. As
$\lambda_{v}>3\sqrt{3}\lambda_{SO}$, the static model describes a
trivial insulator \cite{C. L. Kane1}, then according to the
bulk-edge correspondence, there is no edge state localized at the
open boundary. With the introduction of periodically driving, we
find when the driving frequency $\omega$ is much larger than other
energy scales, there is a large energy gap between Floquet bands
and no edge states traversing the gap.  By decreasing the driving
frequency, we find the gap at $\varepsilon=\pi/T$ will close and
then reopen, with edge states emerging and traversing the reopened
gap as shown in Fig.\ref{fig1}(a), which suggests the Floquet band
now is topologically nontrivial. The edge states are not chiral
since each edge has states which propagate in both directions. As
the Hamiltonian holds the time-reversal symmetry, the edge states
are helical in the sense that fermionic atoms with opposite spin
propagate in opposite direction, therefore, the system now is a
driven QSH insulator. We name such driven QSH insulators as FQSH
insulators. The helical edge states traversing the gap at
$\varepsilon=\pi/T$ is a unique property of driven systems. For a
static system, the helical edge states always appear in the gap at
$\varepsilon=0$ because the spectrum of a static system is bounded
\cite{M. S. Rudner}.

For the sake of discussion, we introduce two $Z_{2}$ topological
indices $\nu_{0}$ and $\nu_{\pi}$ to characterize the topological
properties corresponding to the gap at $\varepsilon=0$ and
$\varepsilon=\pi/T$, respectively. With this introduction, the
trivial phase in the large driving frequency region is
characterized by $(\nu_{0},\nu_{\pi})=(0,0)$ and the FQSH
insulator exhibited in Fig.\ref{fig1}(a) is characterized by
$(\nu_{0},\nu_{\pi})=(0,1)$.

With a further decrease of the driving frequency, the gap at
$\varepsilon=0$ will also close and then  reopen. As a
consequence,  helical edge states traversing both gaps at
$\varepsilon=\pi/T$ and $\varepsilon=0$, and therefore, there are
two pairs of helical states propagating on the same boundary. Such
``anomalous" edge states are without an analog in static system.
In static QSH insulator, when there are even pairs of helical edge
states, the edge states are no longer stable against disorder and
one can always add some extra term to gap all of them, as a
result, the system is topologically equivalent to a trivial
insulator. However, for here the FQSH insulator characterized by
$(\nu_{0},\nu_{\pi})=(1,1)$, the helical edge states traversing
the gaps at $\varepsilon=\pi/T$ and $\varepsilon=0$ are separated
by a big energy difference, as a result, their coupling effects
can be neglected, and the two pairs of helical edge states are
still stable against disorder. Fig.\ref{fig1}(c)(d) show that
these helical states are well localized at the two open boundaries
of the system.

{\it Periodic perturbation breaking the time-reversal symmetry---}
We add a periodic perturbation to the system,
\begin{eqnarray}
V=V_{0}\sigma_{z}\cos(\omega t). \label{8}
\end{eqnarray}
This perturbation breaks the time-reversal symmetry in the sense:
$\mathcal{T}V\mathcal{T}^{-1}=-V$, with
$\mathcal{T}=i\sigma_{y}\mathcal{K}$. Although the form of this
perturbation is the same as the one in Ref.\cite{N. H. Lindner},
the spaces on which the Pauli matrix $\sigma_{z}$ operates are
different. In Ref.\cite{N. H. Lindner},  $\sigma_{z}$ operates on
the subspace of bands (similar to here the subspace of the two
sublattices) in each $2\times2$ block, and therefore, it does not
break the whole system's time-reversal symmetry. That's the reason
why a pair of helical edge states (a pair of chiral edge states in
each $2\times2$ block) can be driven up by the perturbation of
this form. However, $\sigma_{z}$ here is a Pauli matrix purely
operating on spin, consequently, it directly breaks the whole
system's time-reversal symmetry. For such a periodic perturbation,
we find the original time-reversal symmetric static system can not
be driven into a FQSH insulator but the perturbation also does not
affect the edge states of a static QSH insulator and a FQSH
insulator. The robustness of the FQSH insulator against the
time-reversal-symmetry-breaking driving term implies that if there
exist other driving terms which do not break the time-reversal
symmetry, the system can be driven to be a FQSH insulator.

\begin{figure}
\subfigure{\includegraphics[width=4cm, height=4cm]{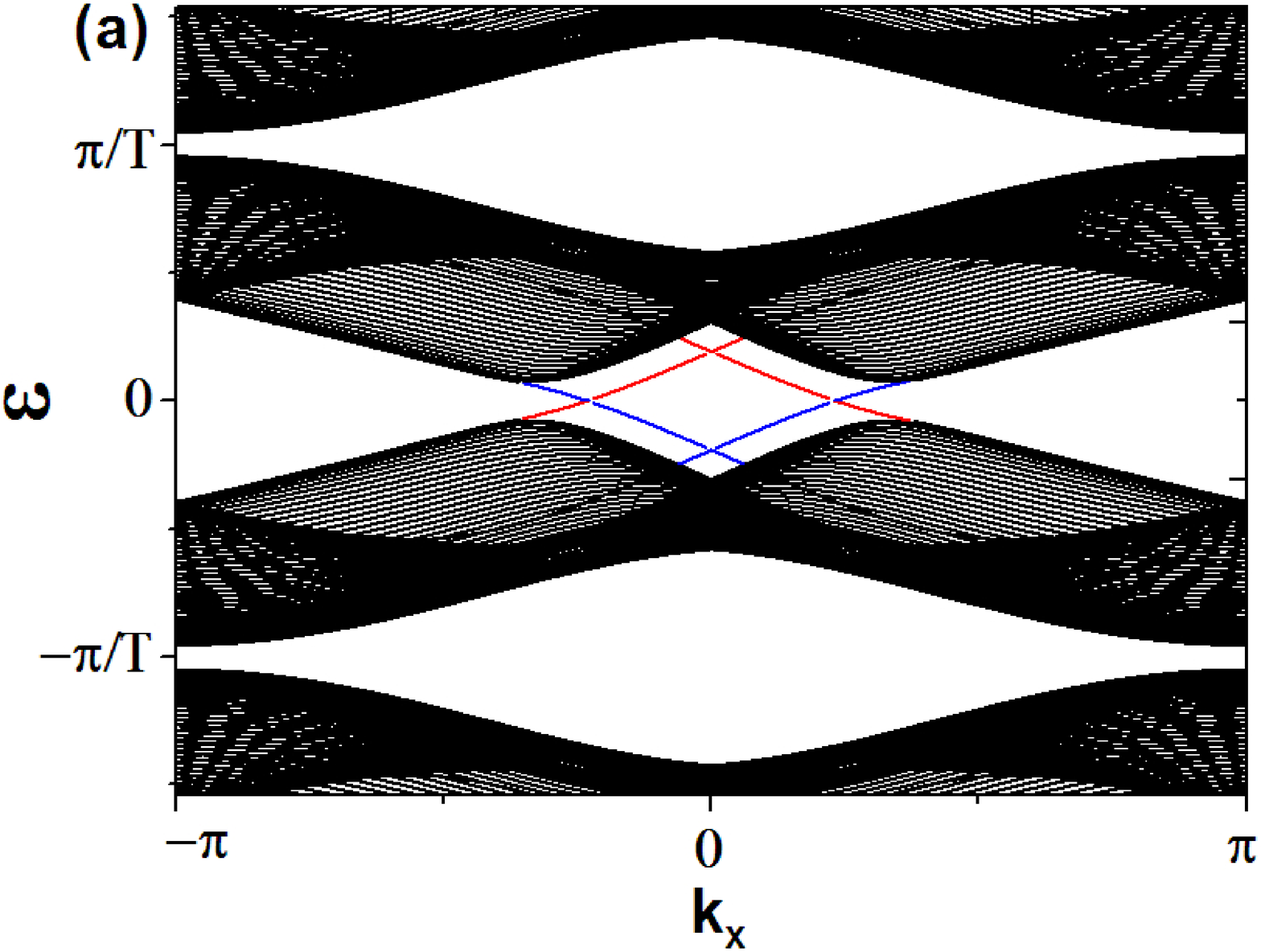}}
\subfigure{\includegraphics[width=4cm, height=4cm]{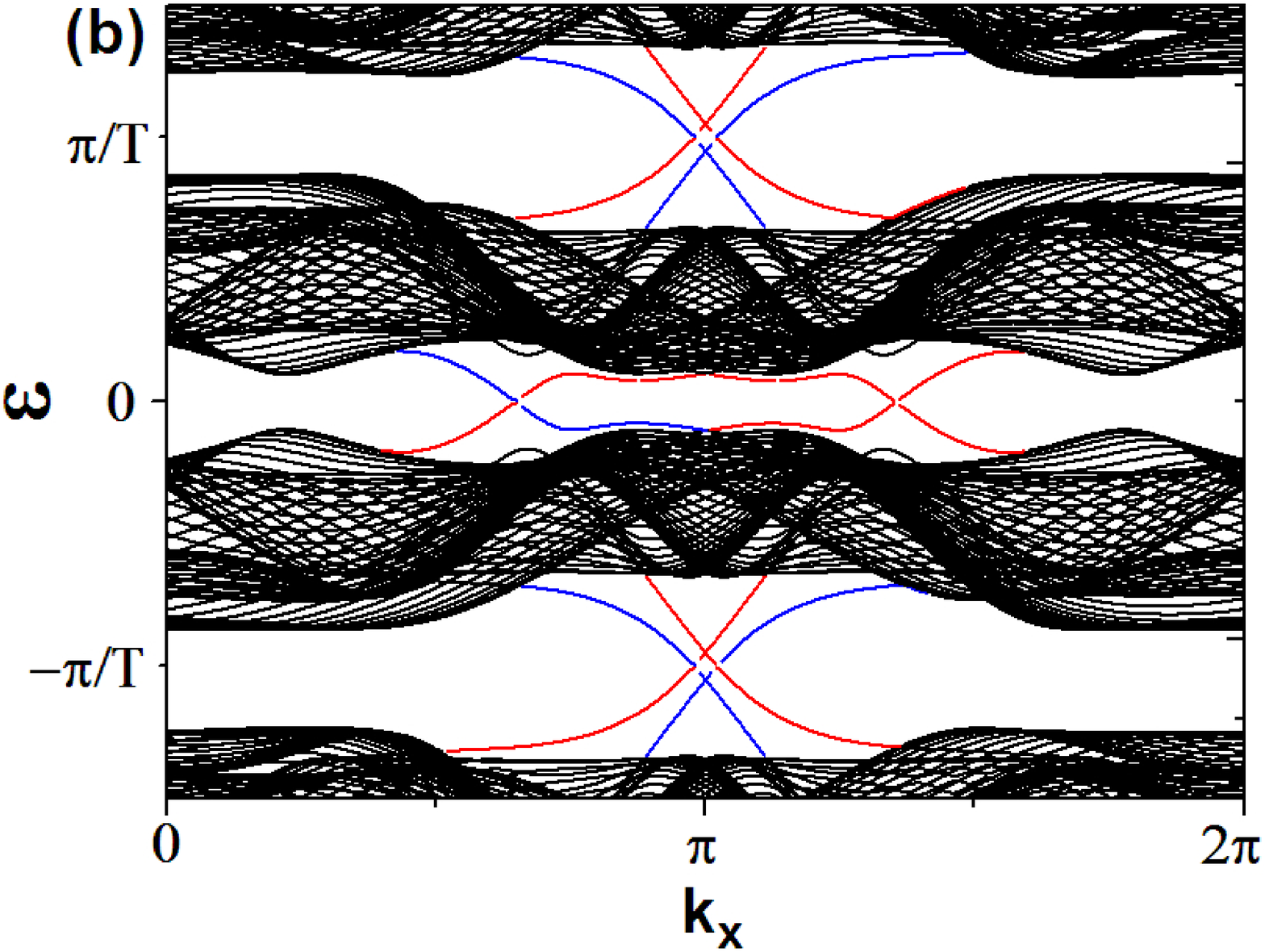}}
\caption{ (color online) (a)(b) Spectrum of the truncated Floquet Hamiltonian (Eq.(\ref{6})) in
the cylindrical geometry.The parameters used in (a) are: $J=1$, $\lambda_{SO}=0.2$,
$\lambda_{v}=0.8$,  $\omega=6.5$, $V_{0}=1$. As $\lambda_{v}<3\sqrt{3}\lambda_{SO}$,
the static system is a QSH insulator, with helical edge states traversing the gap.
The parameters in (b) are: $J=1$, $\lambda_{SO}=0.2$,
$\lambda_{v}=1.2$, $J_{D}=1$, $\lambda_{SO,D}=0.1$, $\omega=2$, $V_{0}=1$.}\label{fig2}
\end{figure}

For details, the static system with the parameters in
Fig.\ref{fig2}(a) is a QSH insulator, with helical edge states
traversing the gap at $\varepsilon=0$. With the introduction of
the time-reversal-symmetry-breaking perturbation, we find no edge
states are driven up in the gap at $\varepsilon=\pi/T$, and the
topology of the system is stable against the perturbation: the
original helical edge states traversing the gap at $\varepsilon=0$
is almost unaffected by the perturbation. The robustness is due to
that the meaning of  ``time-reversal-symmetry-breaking" here is
quite different from the one for static system. In Ref.\cite{N. H.
Lindner}, the authors have discussed that even the Hamiltonian at
any given time may not possess any symmetry under time reversal,
the Floquet Hamiltonian possess the time-reversal symmetry as long
as the condition
$\mathcal{T}\hat{\mathcal{H}}(t)\mathcal{T}^{-1}=\hat{\mathcal{H}}(-t+\tau)$
holds (for some fixed $\tau$). That's the reason why the helical
edge states are not gapped out.

For a FQSH insulator, we find the condition can not be satisfied
due to the existence of three driving terms (two terms from
Eq.(\ref{3})), however, compared Fig.\ref{fig2}(b) to
Fig.\ref{fig1}(b), it is direct to see that the helical edge
states traversing both gaps are also robust against the
perturbation. This may suggest that only the driven terms which do
not hold the time-reversal symmetry  is needed to satisfy the
condition. This is reasonable, when the driven Hamiltonian only
have driving terms which do not break the time-reversal symmetry,
the Hamiltonian is time-reversal invariant at any time,
consequently, the Floquet Hamiltonian is no doubt time-reversal
invariant.

{\it Theoretical model without time-reversal symmetry---} To see
whether such a driving approach can also drive a trivial system
without time-reversal symmetry to be  topological, we consider the
familiar two-band tight-band model realized on a square lattice
\cite{X. L. Qi},
\begin{eqnarray}
H_{0}&=&-\frac{J}{2}\sum_{i}[\psi_{i}^{\dag}i \sigma_{x} \psi_{i+\hat{x}}+\psi_{i}^{\dag}i\sigma_{y}\psi_{i+\hat{y}}+h.c.] \nonumber \\
&+&\frac{B}{2}\sum_{i}[\psi_{i}^{\dag}\sigma_{z}\psi_{i+\hat{x}}+\psi_{i}^{\dag}\sigma_{z}\psi_{i+\hat{y}}+h.c.] \nonumber\\
&+&M\sum_{i}\psi_{i}^{\dag}\sigma_{z}\psi_{i}.\label{9}
\end{eqnarray}
Here $\vec{\sigma}=(\sigma_{x},\sigma_{y},\sigma_{z})$ are Pauli
matrices operating on spin, $J$ denotes the strength of spin-orbit
coupling, $B$ and $M$ denote the difference between the two spin
degrees' hopping amplitude and on-site energy. Without loss of
generality, we assume $B>0$ in this work.

Re-expressing the Hamiltonian under the representation
$\psi_{\mathbf{k}}^{\dag}=(c^{\dag}_{\uparrow,\mathbf{k}},c^{\dag}_{\downarrow,\mathbf{k}})$
in momentum space, we obtain
\begin{eqnarray}
H_{0}=\sum_{\mathbf{k}}\psi_{\mathbf{k}}^{\dag}\mathcal{H}_{0}(\mathbf{k})\psi_{\mathbf{k}},\nonumber
\end{eqnarray}
with
\begin{eqnarray}
\mathcal{H}_{0}(\mathbf{k})&=&\vec{d}(\mathbf{k})\cdot\vec{\sigma}=J\sin(k_{x})\sigma_{x}+J\sin(k_{y})\sigma_{y} \nonumber\\
&&+(M+B(\cos(k_{x}) +\cos(k_{y})))\sigma_{z}.\label{10}
\end{eqnarray}
This Hamiltonian is obviously without time-reversal symmetry, consequently,
when $\vec{d}(\mathbf{k})$ does not vanish in the Brillouin zone,
the topology of this static Hamiltonian is determined by the first Chern number \cite{D. Thouless},
\begin{eqnarray}
C_{1}=\frac{1}{4\pi}\int
d^{2}\mathbf{k}(\frac{\partial\hat{d}(\mathbf{k})}{\partial k_{x}}
\times \frac{\partial\hat{d}(\mathbf{k})}{\partial
k_{y}})\cdot\hat{d}(\mathbf{k}),\label{11}
\end{eqnarray}
where $\hat{d}(\mathbf{k})=
\vec{d}(\mathbf{k})/|\vec{d}(\mathbf{k})|$, and the Chern number
of this system is
\begin{eqnarray}
C_1=\left\{\begin{array}{cc}1,&0<M<2B\\-1,&-2B<M<0\\0,&\text{otherwise}.\end{array}\right.\label{12}
\end{eqnarray}

By varying the optical lattice periodically, the parameters
appearing in Eq.(\ref{9}) will also vary with time periodically,
$J_{x,y}(t)=J+J_{x,y}^{D}\cos(\omega t)$,
$B_{x,y}(t)=B+B_{x,y}^{D} \cos(\omega t)$ and
$M(t)=M+M_{D}\cos(\omega t)$. If we assume the two bands are
close, $J_{x,y}^{D}$ can be much larger than $B_{x,y}^{D}$,
$M_{D}$. Therefore, without loss of generality, we neglect
$B_{x,y}^{D}$, $M_{D}$ for simplicity, then the time-dependent
Hamiltonian is given as
\begin{eqnarray}
\mathcal{H}(\mathbf{k},t)=\mathcal{H}_{0}(k)+\mathcal{H}_{D}(\mathbf{k})\cos(\omega t),
\end{eqnarray}
where $\mathcal{H}_{D}(\mathbf{k})=J_{x}^{D}\sin(k_{x})\sigma_{x}+J_{y}^{D}\sin(k_{y})\sigma_{y}$.
The effect of varying the optical lattice potential is equivalent to varying
the spin-orbit coupling periodically.

For the isotropic driving case, $J_{x}^{D}$ is equal to $J_{y}^{D}$.
To see how the driving affects the topology of the system, we also
consider the system with periodical boundary condition in $x$ direction
and open boundary condition in $y$ direction.

\begin{figure}
\subfigure{\includegraphics[width=4cm, height=4cm]{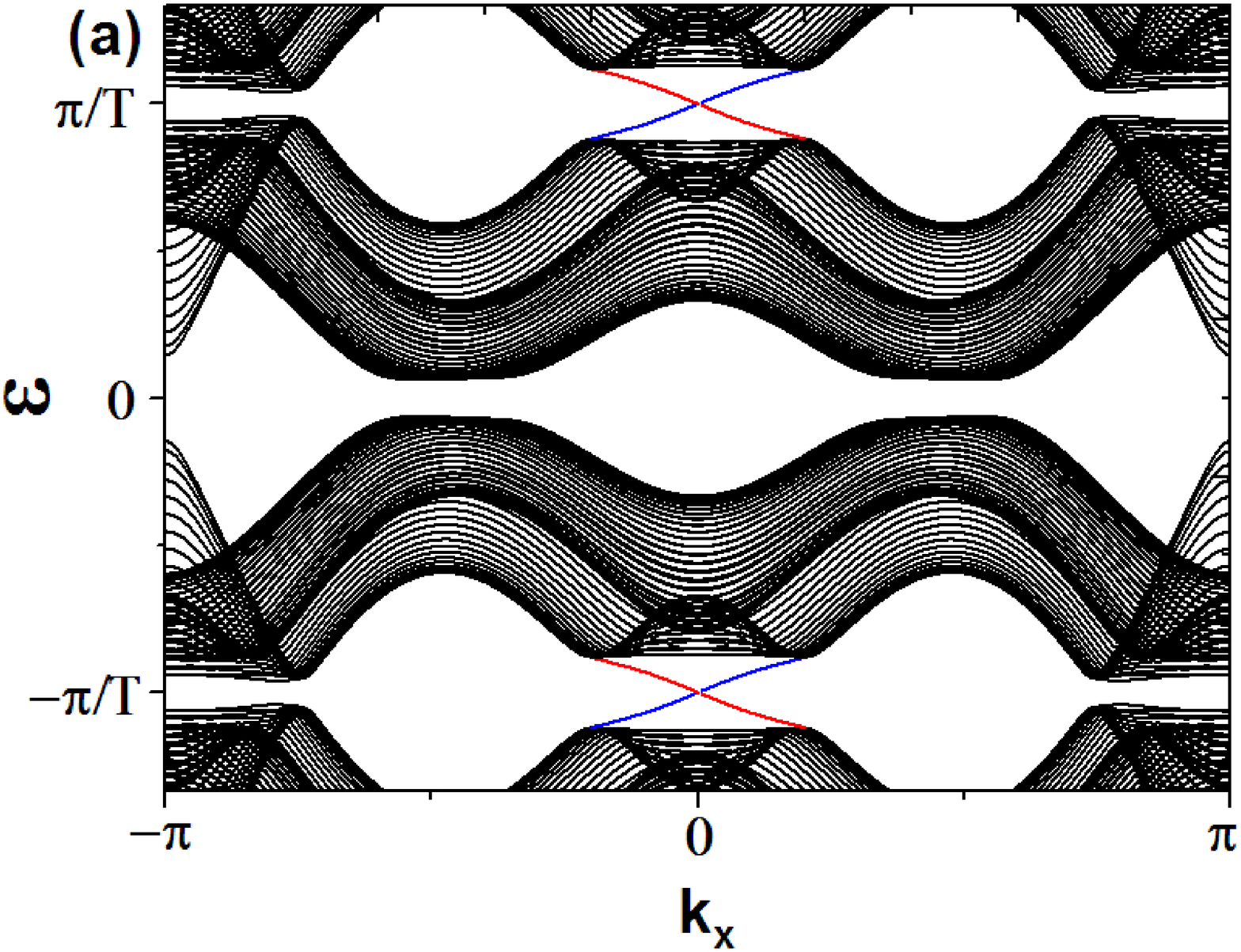}}
\subfigure{\includegraphics[width=4cm, height=4cm]{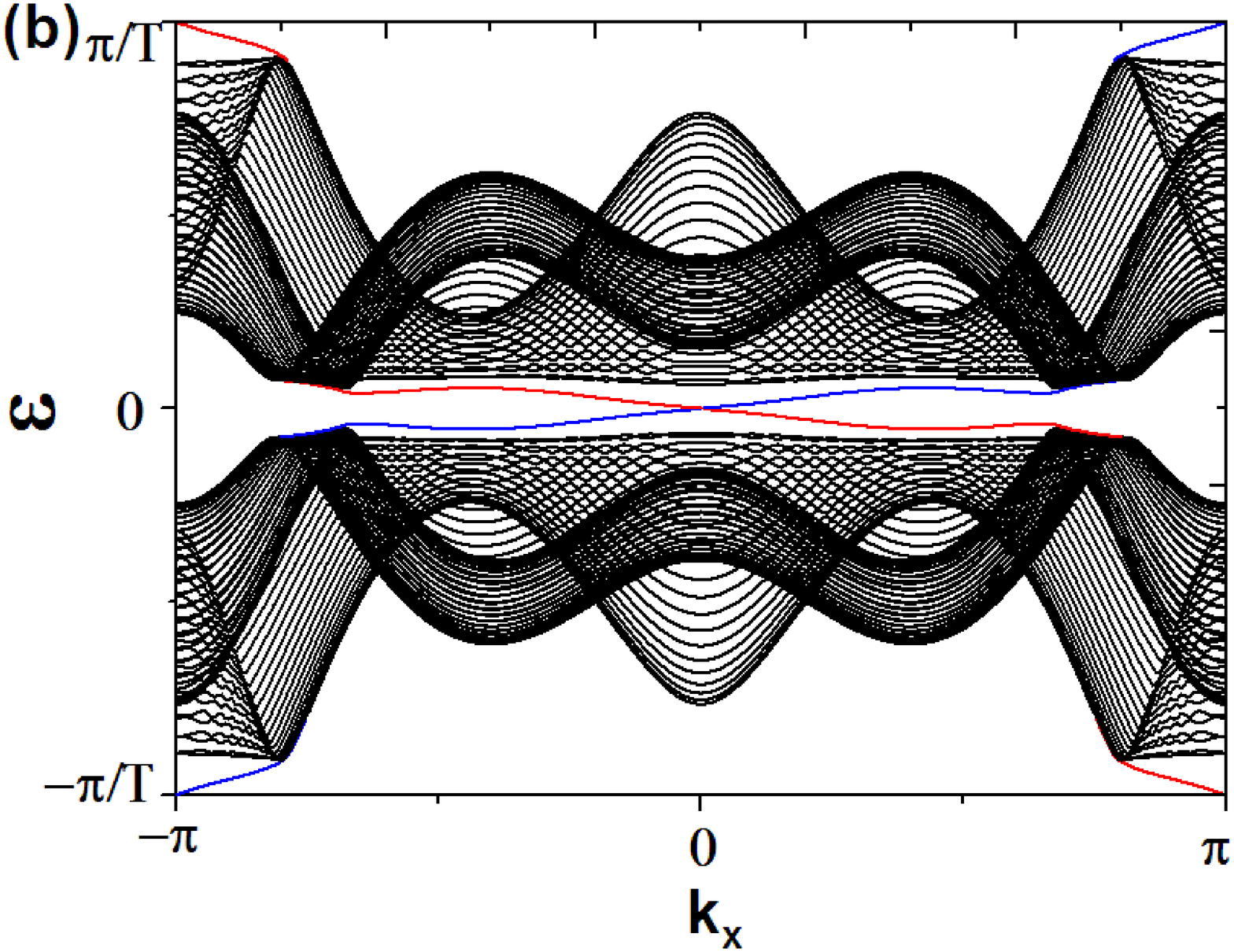}}
\subfigure{\includegraphics[width=4cm, height=4cm]{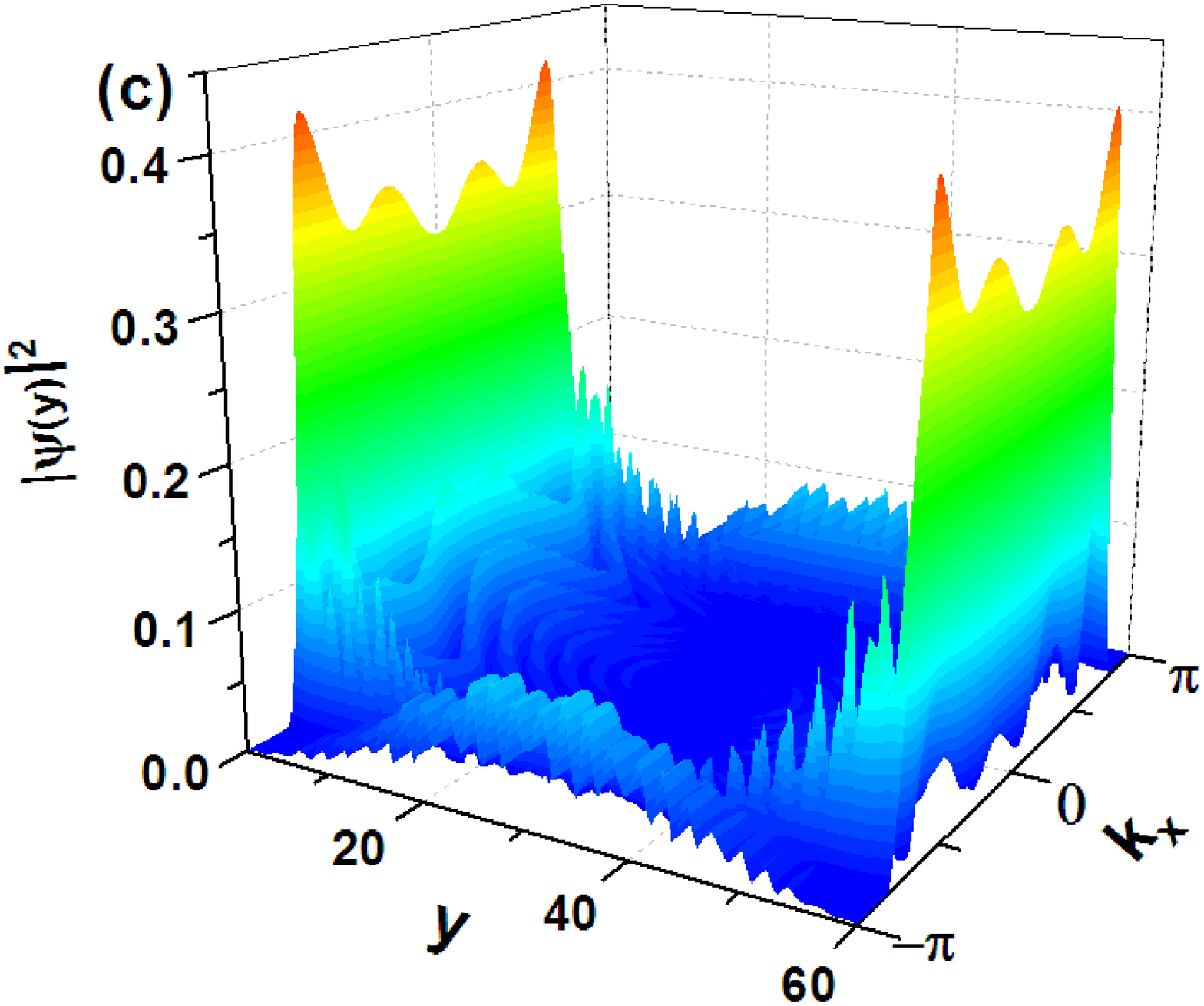}}
\subfigure{\includegraphics[width=4cm, height=4cm]{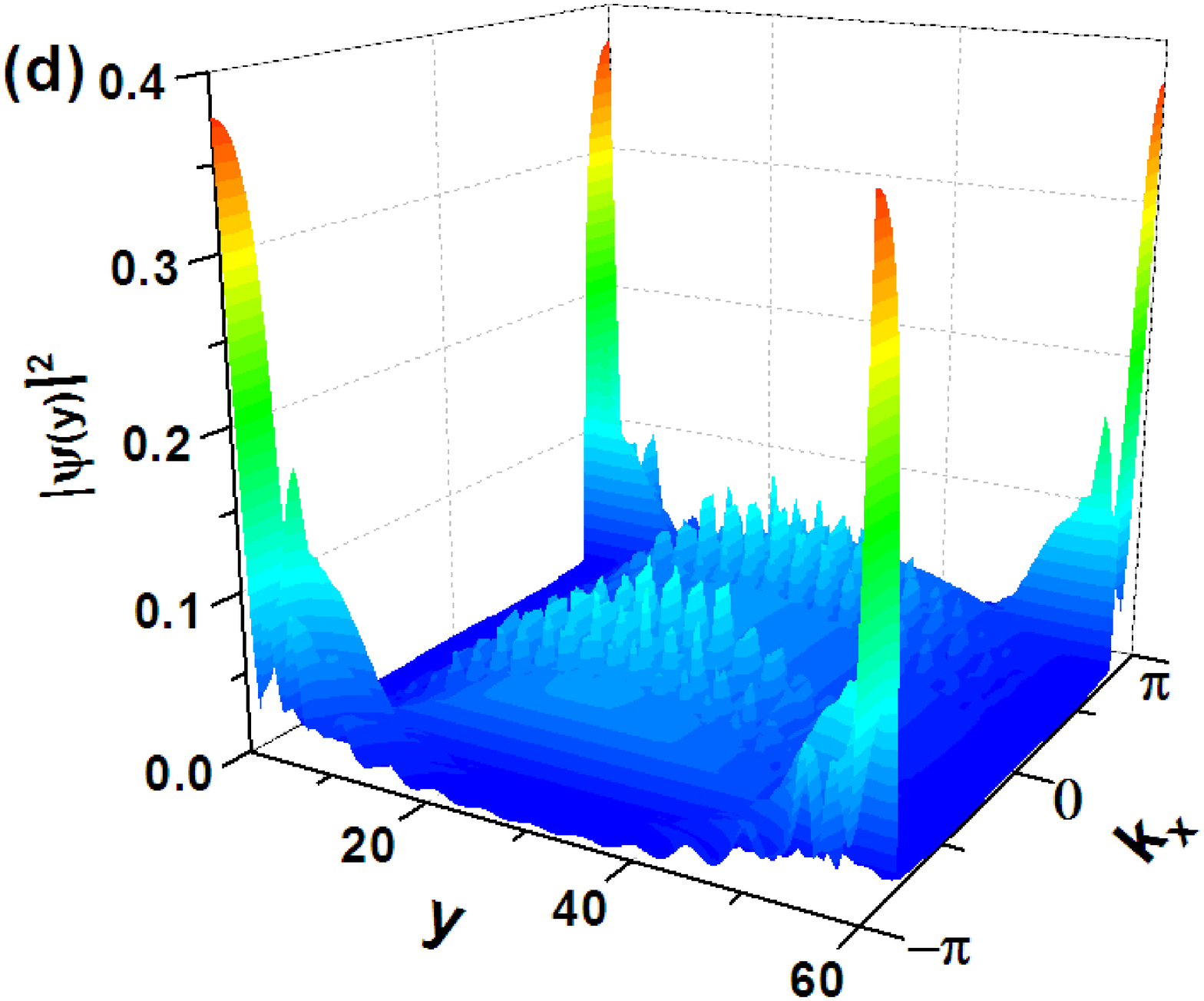}}
\caption{ (color online) (a) (a) Floquet spectrum with parameters: $J$=1, $B=0.2$, $M=0.5$,
$J^{D}_{x,y}=-0.5$, and $\omega=1.5$.
(b) Floquet spectrum with parameters: $J=1$, $B=0.3$, $M=0.8$, $J^{D}_{x,y}=-0.5$, $\omega=1.3$.
(c) The density of edge states corresponding to the gap at the quasienergy $\varepsilon=0$.
(d) The density of edge
states corresponding to the gap at the quasienergy $\varepsilon=\pi/T.$}\label{fig3}
\end{figure}

Based on Eq.(\ref{12}), the parameters in Fig.\ref{fig3}(a)
suggest the system in static is a trivial insulator without edge
states. With the introduction of periodically driving, we find the
picture is similar to the time-reversal symmetric case that when
the driving frequency $\omega$ is much larger than other energy
scales, there is a large energy gap between Floquet bands and no
edge states. With decreasing the driving frequency, the gap at
$\varepsilon=\pi/T$ firstly closes and then reopens, with edge
states emerging and traversing the reopened gap as shown in
Fig.\ref{fig3}(a). As the system is without time-reversal
symmetry, here the edge states are chiral in the sense that the
fermionic atoms with opposite velocity propagate on the opposite
boundary.

With a further decrease of the driving frequency, the gap at
$\varepsilon=0$ will also close and then reopen. As a consequence,
chiral edge states traverse both gaps at $\varepsilon=0$ and
$\varepsilon=\pi/T$. As the winding number of a band is equal to
the difference between the number of edge states at the gaps above
and below the band,
$C_{\varepsilon\varepsilon^{'}}=n_{edge}(\varepsilon)-n_{edge}(\varepsilon^{'})$,
it is direct to see that the two bands' winding numbers in
Fig.\ref{fig3}(c) are both {\it zero}. This is another unique
property of a periodically driving system that the chiral edge
states can exist despite the fact that the Chern numbers
associated with both bands are zero \cite{M. S. Rudner}.  These
chiral states are localized at the two open boundaries of the
system, as shown in Fig.\ref{fig3}(c)(d).

\begin{figure}
\subfigure{\includegraphics[width=4cm, height=4cm]{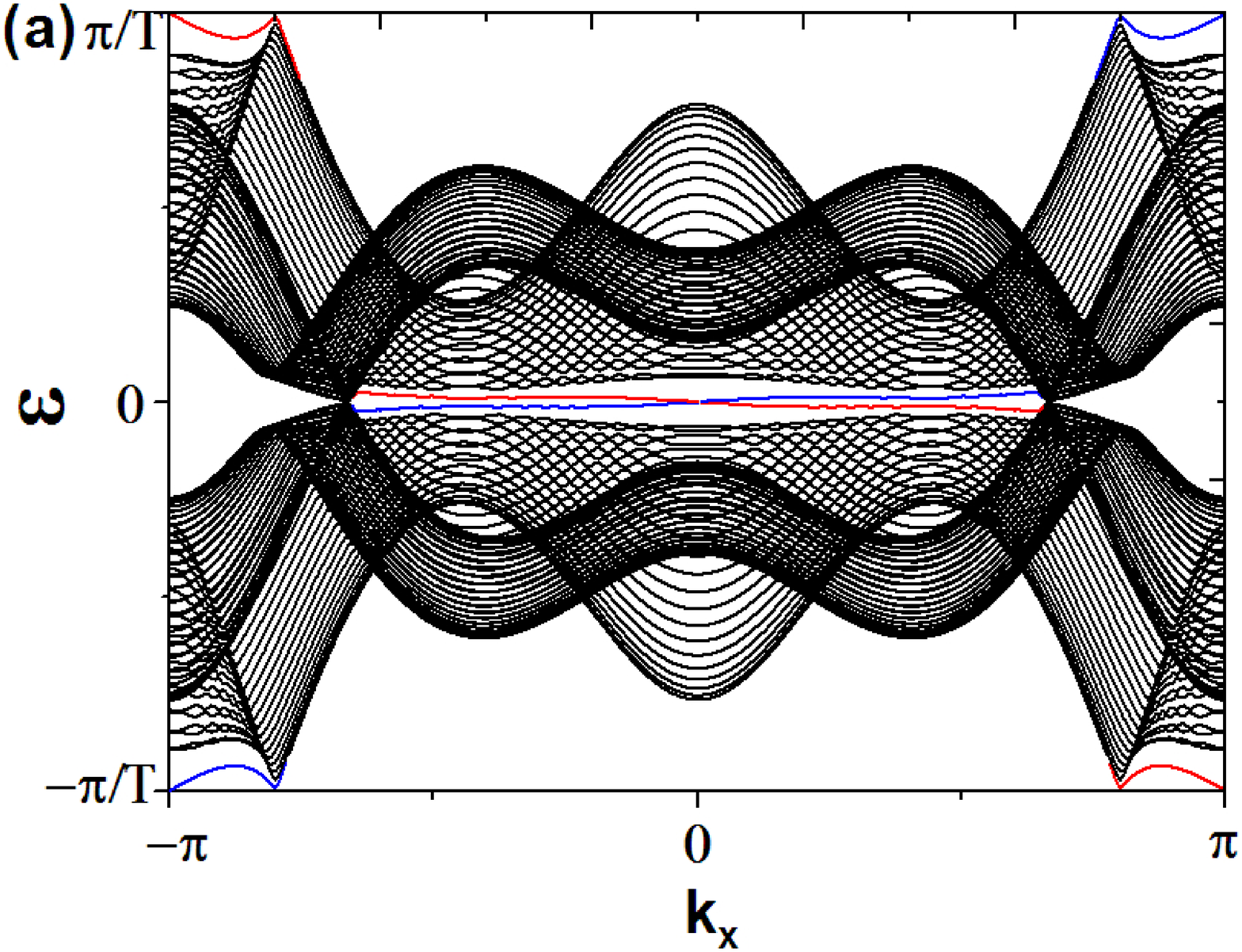}}
\subfigure{\includegraphics[width=4cm, height=4cm]{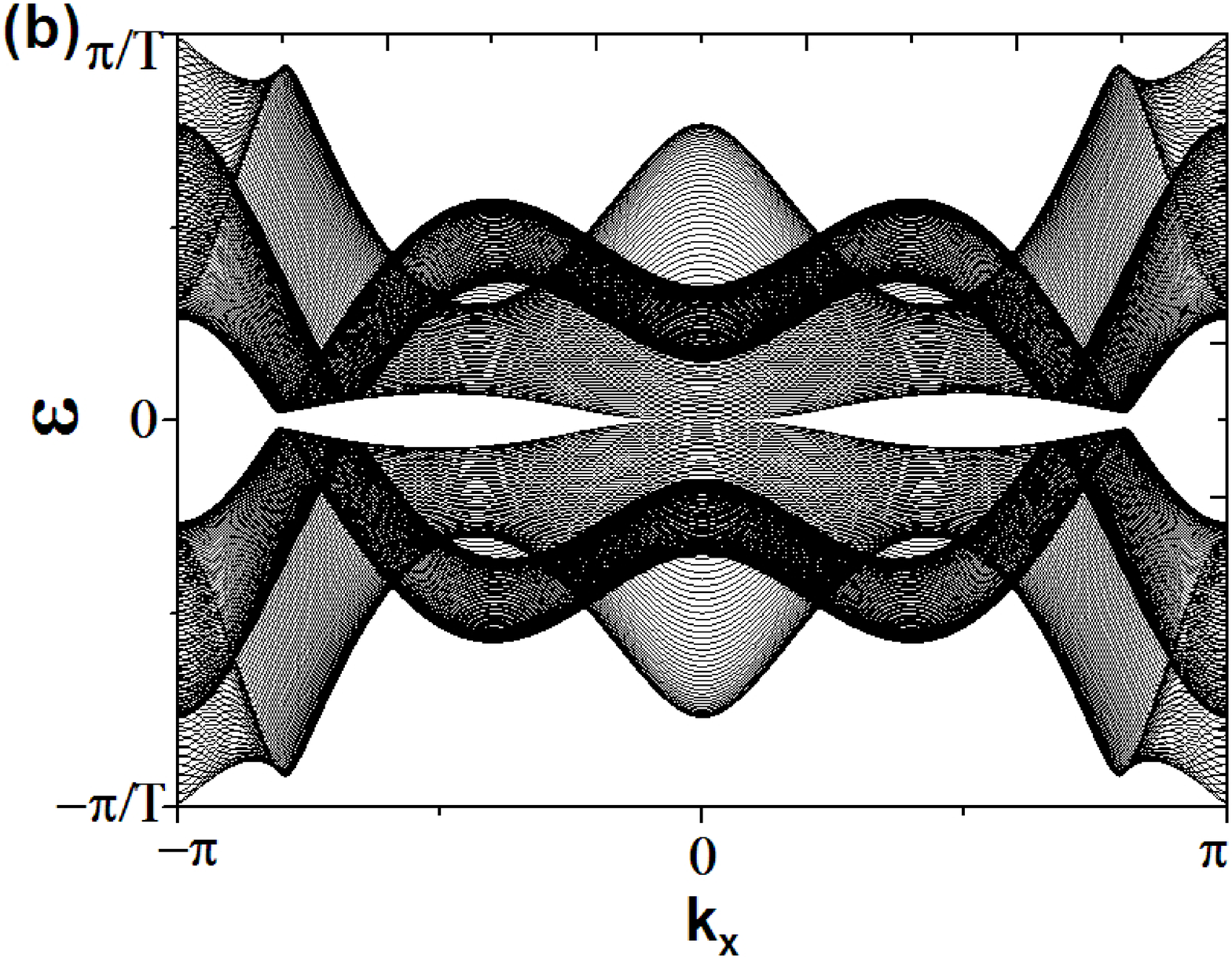}}
\caption{ (color online) Anisotropic driving cases. (a) Driving along the $y$ direction,
Floquet spectrum with parameters: $J$=1, $B=0.2$, $M=0.5$,
$J^{D}_{x}=0$, $J^{D}_{y}=-0.5$, and $\omega=1.3$.
(b) Driving along the $x$ direction, Floquet spectrum with parameters:
$J=1$, $B=0.3$, $M=0.8$, $J^{D}_{x}=-0.5$, $J^{D}_{y}=0$ and $\omega=1.3$.
}\label{fig4}
\end{figure}

If we only drive the system along the $y$ direction, $i.e.$
$J_{D,x}=0$, $J_{D,y}\neq0$, we find the results are similar to
the isotropic case's. Fig.\ref{fig4}(a) shows that chiral edge
states traverse both gaps, however, compared to the isotropic case
under the same parameter condition except $J_{x}^{D}$, we find the
gaps at $\varepsilon=0$ and $\varepsilon=\pi/T$ are greatly
decreased. If we instead only drive the system in the $x$
direction, in other word, $J_{D,x}\neq0$, $J_{D,y}=0$, the picture
is dramatically changed. No matter what parameters are chosen,
there is no edge state emerging, therefore, the system is always a
trivial insulator, as shown in Fig.\ref{fig4}(d). Although driving
the system along the direction with periodical boundary condition
will not induce chiral edge states at the open boundary, such a
driving has the effect that it enlarges the energy gap.

{\it  Conclusions---} We find that the optical lattice potential
vary periodically provides a simple, general, and realizable way
to drive a cold atomic system with or without time-reversal
symmetry into phases with nontrivial topology. For a
$\mathcal{T}$-invariant system, we find that this simple approach
can drive the trivial insulator into a FQSH insulator but an
external time-dependent field which couples with the spin and
consequently breaks the time-reversal symmetry can not. The FQSH
insulator, a novel state similar to the QSH insulator, can host
one or two pair of helical edge states at the same boundary, and
the edge states are robust against the
time-reversal-symmetry-breaking periodic perturbation. Applying
this driving approach to a system without time-reversal symmetry,
we find that edge states driven up are chiral and the picture is
similar to the one by driving the system with an external
electromagnetic field.

{\it  Acknowledgments---} This work was supported by NSFC Grant
No.11275180 and National Science Fund for Fostering Talents in
Basic Science No.J1103207.

\end{document}